\documentclass{aa}
\usepackage{txfonts}
\usepackage{epsfig}
\usepackage{subfigure}
\usepackage{multirow}

\usepackage{times}
\usepackage{natbib}
\bibpunct{(}{)}{;}{a}{}{,}

\begin{document}

\title{A search for changing-look AGN in the Grossan catalog}
\author{S. Bianchi\inst{1}, M. Guainazzi\inst{1}, G. Matt\inst{2}, M. Chiaberge\inst{3,4}, K. Iwasawa\inst{5}, F. Fiore\inst{6}, R. Maiolino\inst{7}}

\offprints{Stefano Bianchi\\ \email{Stefano.Bianchi@sciops.esa.int}}

\institute{XMM-Newton Science Operations Center, European Space Astronomy Center, ESA, Apartado 50727, E-28080 Madrid, Spain
\and Dipartimento di Fisica, Universit\`a degli Studi Roma Tre, Via della Vasca Navale 84, I-00146, Roma, Italy
\and Space Telescope Science Institute, 3700 San Martin Drive, Baltimore, MD 21218
\and INAF-Istituto di Radioastronomia, Via P. Gobetti 101, 40129 Bologna, Italy
\and Institute of Astronomy, Madingley Road, Cambridge, CB3 0HA
\and INAF - Osservatorio Astronomico di Roma, via di Frascati 33, I-00040 Monteporzio, Italy
\and INAF - Osservatorio Astrofisico di Arcetri, Largo Fermi 5, I-50125 Firenze, Italy}

\date{Received / Accepted}

\authorrunning{S. Bianchi et al.}

\abstract{We observed with XMM-\textit{Newton} 4 objects selected from the \citet{grossan92} catalog, with the aim to search for new `changing-look' AGN. The sample includes all the sources which showed in subsequent observations a flux much lower than the one measured with HEAO A-1: NGC~7674, NGC~4968, IRAS~13218+0552 and NGC~1667. None of the sources was caught in a high flux state during the XMM-\textit{Newton} observations, whose analysis reveal they are all likely Compton-thick objects. We suggest that, for all the sources, potential problems with the HEAO A-1 source identification and flux measurement prevent us from being certain that the HEAO A-1 data represent a putative `high' state for these objects. Nonetheless, based on the high flux state and Compton-thin spectrum of its \textit{GINGA} observation, NGC~7674 represents probably the sixth known case of a `changing-look' Seyfert 2 galaxy. From the X-ray variability pattern, we can estimate a likely lower limit of a few parsec to the distance of the inner walls of the torus in this object. Remarkably, IRAS~13218+0552 was not detected by XMM-\textit{Newton}, despite being currently classified as a Seyfert 1 with a large [OIII] flux. However, the original classification was likely to be affected by an extreme velocity outflow component in the emission lines. The object likely harbors an highly obscured AGN and should be re-classified as a Type 2 source.

\keywords{galaxies: Seyfert - X-rays: galaxies - X-rays: individual: NGC7674 - X-rays: individual: NGC1667 - X-rays: individual: NGC4968 - X-rays: individual: IRAS13218+0552}

}

\maketitle

\section{Introduction}

Recently, a few Seyfert 2 galaxies were discovered to make transitions, on time scales of a few years, from a Compton--thin appearance (when the nuclear radiation is absorbed by material with a line--of--sight column density less than $\sigma_T^{-1}$=1.5$\times10^{24}$ cm$^{-2}$) to a reflection--dominated spectrum, and/or viceversa \citep{gua02, gua02b, mgm03}. A reflection--dominated spectrum is recognized by a hard continuum and a prominent (Equivalent Width, EW, $\sim$1 keV) iron line, and it is commonly assumed to be a signature of Compton--thick
absorption \citep{matt00b, matt02b}. In this hypothesis, the reflection would be due to circumnuclear material, including the part of the absorbing matter which is visible to both the nucleus and the observer, such as the far side of the inner wall of the torus envisaged in Unification Models \citep{antonucci93}.

Two possibilities exist for the observed transitions: either a change in the line-of-sight absorbing column, or a `switching-off' of the nucleus, leaving reflection from distant matter as an echo of past activity. While the first hypothesis cannot in most cases be ruled out \citep[and, indeed, seems to be the best explanation in the case of NGC~1365:][]{ris05}, the second appears more likely \citep[see discussion in ][]{mgm03} and in at least one case, namely NGC~2992 \citep{gilli00}, it is the only tenable.
These sources would then be the Seyfert 2 analogs of the Narrow Line Seyfert 1 NGC~4051 \citep{gua98,utt99}. In the following, we will refer to all sources (whether Seyfert 1s or 2s) with the term `changing look', if they temporarily appear to be reflection--dominated.

These objects have been serendipitously discovered; the very nature of this transition makes a systematic study of their properties difficult. In particular, how frequently sources undergo this sort of transition? Moreover, how many Compton--thick Seyfert 2 are really heavily absorbed, and not simply switched--off? It is quite difficult to estimate the fraction of these transitions,
due to the lack of a complete and unbiased sample of homogeneously defined Seyfert galaxies with sufficient X-ray
temporal and spectroscopic coverage. A first step in this direction was done by \citet{gua05} who reported a typical occurrence rate of a transition every 50 years, on the basis of a sample of 11 optically-selected Seyfert 2 galaxies, whose \textit{ASCA} and/or BeppoSAX observations suggested Compton-thick obscuration. In this paper, we select a mini-sample of other four `changing-look' candidates with long-term X-ray coverage.

\section{Observations}

\subsection{The sample}

Our sample is extracted from the \citet{grossan92} catalog, based on the catalog of X-ray sources from the LASS (Large Area Sky Survey, or HEAO A-1) instrument aboard the HEAO-1 satellite, which included the first scan of the full sky from August 1977 to February 1978 \citep{wood84}. The catalog took advantage of the overlapping diamond-shaped error regions of the MC instrument \citep[Modulation Collimator, or A-3 experiment:][]{gursky78}, which also flew aboard the HEAO-1 satellite, to identify the 96 AGN that make up the so-called LMA (LASS/MC identified AGN) sample, down to a limiting 2-10 keV flux of about 1.8$\times10^{-11}$ erg cm$^{-2}$ s$^{-1}$.

Our sample selection criteria aimed at finding good candidates for `changing--look' sources, to be observed with XMM-\textit{Newton}. The selection was made by taking all the sources in the Grossan catalog which have subsequently been observed in X--rays to be reflection--dominated, or too faint to allow for a detailed spectral analysis, with a flux at least a factor 50 fainter than observed by HEAO A-1. If the optical identifications are correct, these sources could therefore be `changing-look' sources.

These selection criteria resulted in a mini-sample of four targets: NGC~7674, NGC~4968, IRAS~13218+0552 and NGC~1667. We note here that none of these sources belongs to the \citet{pic82} HEAO A-2 sample, because they have all 2-10 keV fluxes below its limit.

\begin{table*}
\caption{\label{sample}The sources included in our sample and the log of the XMM-\textit{Newton} observations.}
\begin{center}
\begin{tabular}{ccccccccc}
\hline
~ & ~ & ~ & ~ & ~ & ~ & ~ &\cr
\textbf{Name} & \textbf{LASS Name} & \textbf{z} & \textbf{Type (NED)} & \textbf{[OIII]$^a$} & \textbf{IR$^b$} & \textbf{XMM Obs. Date} & \textbf{ObsID} & \textbf{Exp. (ks)}\cr
~ & ~ & ~ & ~ & ~ & ~ & ~ & ~ &\cr
\hline
~ & ~ & ~ & ~ & ~ & ~ & ~ & ~ &\cr
NGC~7674 & 1H2320+084 & 0.029 & Sey~2 & 1.9 & 4.7 & 2004-06-02 & 0200660101 & 9\cr
~ & ~ & ~ & ~ & ~ & ~ & ~ & ~ &\cr
\multirow{2}*{NGC~4968} & \multirow{2}*{1H1308-237} & \multirow{2}*{0.01} & \multirow{2}*{Sey~2} & \multirow{2}*{11.1} & \multirow{2}*{2.7} & 2001-01-05 & 0002940101 & 4\cr
~ & ~ &~ & ~ & & &2004-07-05 & 0200660201 & 10 \cr
~ & ~ &~ & ~ & ~ & ~ & ~ & ~ &\cr
IRAS~13218+0552 & 1H1320+066 & 0.21 & Sey~1 & 0.73 & 1.1 & 2004-07-11 & 0200660301 & 10\cr
~ & ~ & ~ & ~ & ~ &~ & ~ & ~ &\cr
NGC~1667 & 1H0445-060 & 0.015 & Sey~2 & 2.0 & 3.8 & 2004-09-20 & 0200660401 & 8\cr
~ & ~ & ~ & ~ & ~ &~ & ~ & ~ &\cr
\hline
\end{tabular}
\end{center}
$^a$ [OIII] fluxes in units of $10^{-12}$ erg cm$^{-2}$ s$^{-1}$, corrected for extinction adopting the prescription given by \citet{bass99}. All data were taken from \citet{polletta96}, except for IRAS~13218+0552 \citep{kvs98}. - $^b$ Far-infrared fluxes in units of $10^{-10}$ erg cm$^{-2}$ s$^{-1}$, defined as F$_{IR}$~=~F$_{25\mu m}$~x~$\nu_{25\mu m}$~+~F$_{60\mu m}$~x~$\nu_{60\mu m}$. All data were taken from \citet{polletta96}, except for IRAS~13218+0552 \citep{ks98}.
\end{table*}

\subsection{Data reduction}

Table \ref{sample} shows the log of all the XMM-\textit{Newton} observations analysed in this paper. All observations were performed with the EPIC CCD cameras, the pn \citep{struder01} and the two MOS \citep{turner01}, but with different combinations of subframes and filters. Data were reduced with \textsc{SAS} 6.1.0 \citep{sas610} and screening for intervals of flaring particle background was done consistently with the choice of extraction radii, in an iterative process based on the procedure to maximize the signal-to-noise ratio described by \citet{pico04}. No source suffers from pileup problems, so pn spectra were extracted with pattern 0 to 4 and MOS spectra with patterns 0 to 12. When the two MOS observations were performed with the same subframe, their spectra were summed. All spectra were binned in order to oversample the instrumental resolution by at least a factor of 3 and to have no less than 25 counts in each background-subtracted spectral channel. The latter requirement allows us to use the $\chi^2$ statistics. On the other hand, `local fits' were also performed in the 5.25-7.25 keV energy range with the unbinned spectra, using the \citet{cash76} statistics, in order to better assess the nature or the presence of the iron lines, like in the cases of NGC~1667 and NGC~7674. We refer the reader to \citet{gua05b} for details on this kind of analysis. However, all the EWs reported in the paper refer to the global fits on the binned spectra and with respect to the best fit model.

For two sources, namely NGC~7674 and IRAS~13218+0552, we also analysed previous BeppoSAX observations, in the latter case for the first time, to better compare them to the XMM-\textit{Newton} results. Event files and spectra were retrieved from the ASDC Multi-Mission Interactive Archive\footnote{http://www.asdc.asi.it/}. MECS spectra for NGC~7674 were also re-extracted from circular regions with different radii, using \textsc{Xselect}. As for the \textit{ASCA} observations of NGC~4968 and NGC~1667, since they are characterised by detections at the limit of the instrument and different fluxes are reported in literature, we re-analyzed archival data, starting from linearized event lists extracted from the HEASARC archive\footnote{http://heasarc.gsfc.nasa.gov/db-perl/W3Browse/w3browse.pl}.

\subsubsection{\label{ginga}A note on \textit{GINGA} observations}

Since the \textit{GINGA} observation of NGC~7674 will be of fundamental importance to understand the nature of this source, we need to consider with some detail the reliability of \textit{GINGA} fluxes of faint objects.
The brightness of the X-ray background (XRB) is known to fluctuate on the sky. With the beam of the \textit{GINGA} LAC ($FWHM = 1^{\circ}\times 2^{\circ}$), $3\sigma$ fluctuation of the XRB was estimated to be $\sim 2$ ct s$^{-1}$ in the 2-10 keV band \citep{hay89,but97}.
Therefore, in the absence of a reliable estimate of the local XRB brightness level, the cosmic variance of the XRB
limits the source detection with a pointed \textit{GINGA} observation.

For some sources, scanning observations were made just before or after pointing at the target object. This will secure the source identification at least in the direction of the scan path within 0.2 degree or less, and reduce the uncertainty due to the cosmic variance of the XRB brightness.

There is a scanning observation for NGC~7674, in which a significant excess was detected at the position of the galaxy, with a count rate of 3.3 ct s$^{-1}$ \citep{awakiphd}. The XRB level was estimated based on the scanning data in \citet{awakiphd}. On the other hand, \citet{sd96} did not consider NGC~7674 to be detected because the source count rate they estimated was smaller than the cosmic variance (1.8 ct s$^{-1}$). \citet{awakiphd} and \citet{sd96} used different methods in estimating the detector background. This might introduce some difference in their measure of the source flux. However, we consider the source detection reported in \citet{awakiphd} to be reliable, as it was based on the scanning observation, which is free from the uncertainty due to cosmic variance in the XRB.

On the other hand, no scanning observation was carried out for NGC~1667 while the local XRB was estimated only based on the data taken at the nearby off-source sky. The source count rate for NGC~1667 reported in \citet{ak93} is 0.5 ct s$^{-1}$, which is smaller than the cosmic variance of the XRB. Because of the lack of a scanning observation,
this small excess recorded for NGC~1667 cannot be considered as a reliable detection and we take this value as an upper limit. An earlier detection reported in \citet{polletta96} came from a detection of a source during a slewing operation of the satellite. The detection is only in the soft band (2--4 keV). An inspection of the \textit{ROSAT} All-Sky Survey (RASS) image finds brighter soft X-ray sources: one at 15 arcmin to the E and two in 75 arcmin to the N and SE, and they are likely to be confused with NGC~1667.\\

All spectra were analyzed with \textsc{Xspec} 11.3.1. In the following, errors correspond to the 90\% confidence level for one interesting parameter ($\Delta \chi^2 =2.71$), where not otherwise stated. The cosmological parameters used throughout this paper are $H_0=70$ km s$^{-1}$ Mpc$^{-1}$, $\Lambda_0=0.73$ and $q_0=0$.

\section{\label{analysis}Analysis}

\subsection{\label{7674}NGC~7674}

NGC~7674 is a Seyfert 2 galaxy with broad H$\alpha$ and H$\beta$ components in polarized light \citep{mg90,young96}. After its LASS 2-10 keV flux of $2.4\times10^{-11}$ erg cm$^{-2}$ s$^{-1}$, \citet{awaki91} reported a flux of $8\times10^{-12}$ erg cm$^{-2}$ s$^{-1}$ with \textit{GINGA}, with a spectral shape characterized by an absorbed powerlaw (but the derived column density was unconstrained) and an upper limit of 80 eV to the iron line EW. Subsequently, a \textit{ROSAT} PSPC spectrum derived a 0.5-2 keV flux of $2\times10^{-13}$ erg cm$^{-2}$ s$^{-1}$ \citep{lwh01}. The source was then observed by BeppoSAX on November 1996, and found clearly reflection--dominated, with a flux of $5\times10^{-13}$ erg cm$^{-2}$ s$^{-1}$ \citep{mal98}. If the source was absorbed, instead of switched--off as suggested by the HEAO A-1 and \textit{GINGA} measurements, the lack of strong excess emission in the PDS instrument permits to put a lower limit to the column density of the absorber of several times 10$^{24}$ cm$^{-2}$.

\begin{figure}
\epsfig{file=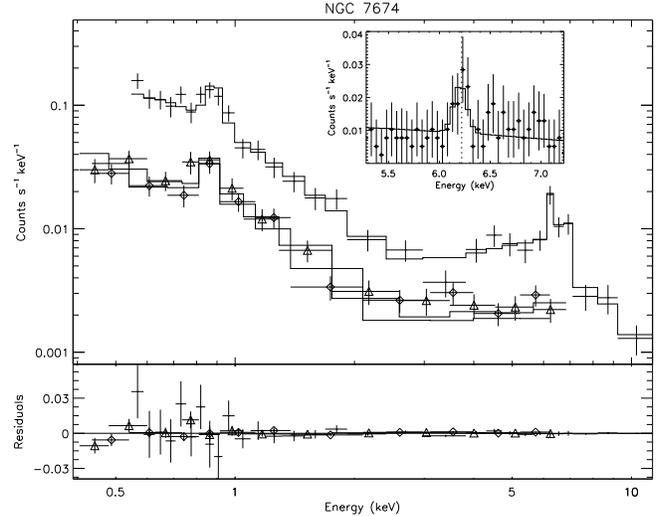, width=8.5cm}
\caption{\label{7674spectrum}NGC~7674: XMM-\textit{Newton} spectra (\textit{pn}: crosses, \textit{MOS1}: diamonds, \textit{MOS2}: triangles) and best fit model. In the inset: `local fit' on the background-subtracted, unbinned pn spectrum around the rest-frame iron line energy, marked with a broken line (see text for details).}
\end{figure}

XMM-\textit{Newton} confirmed the spectral shape observed by BeppoSAX, which we modeled with a bare Compton reflection component \citep[model \textsc{pexrav} in \textsc{Xspec}: ][]{mz95}, a neutral iron K$\alpha$ line and a steep power law for the soft excess, with photon index $\Gamma_s$. Two further emission lines are required by the data, at $0.91^{+0.02}_{-0.03}$ keV ({Ne\,\textsc{ix}} K$\alpha$) and at $6.97^{+0.26}_{-0.05}$ keV, at the 99.9\% and 98\% confidence level, respectively, according to F-test. The latter confirms the complex iron line profile observed with BeppoSAX, but is not unambiguously present in the `local fit' (see inset in Fig. \ref{7674spectrum}): a likely explanation is that the feature is not dominated by a single emission line, but is instead a blend of lines, such as the {Fe\,\textsc{xxvi}} K$\alpha$ and the neutral Fe K$\beta$.

The only problem with a Compton-reflection dominated scenario is the EW of the neutral iron line ($\simeq400$ eV with respect to the reflection component only), which is much lower than the expected one, i.e $>\sim1$ keV \citep[see e.g. ][]{matt96}. It is likely that the measure of the flux of the iron line in NGC~7674 is affected by the above-mentioned prominent feature at higher energies, considering that the EPIC spectra do not have high statistics. In any case, EW as low as $\simeq600$ eV have been measured in Compton-thick sources and may be due to iron underabundance and/or a small inclination angle of the torus \citep[see e.g. Mrk~3: ][]{bianchi05b}.

The XMM-\textit{Newton} spectrum is plotted in Fig. \ref{7674spectrum}, while Table \ref{7674table} summarizes the best fit parameters compared to our re-analysis of the BeppoSAX data. No significant variability is found in any of the spectral parameters. Moreover, we find a somewhat larger BeppoSAX 2-10 keV flux with respect to the one reported by \citet{mal98} (but they do not quote any error), in better agreement with the $7\times10^{-13}$ erg cm$^{-2}$ s$^{-1}$ measured with XMM-\textit{Newton}. We note that the BeppoSAX flux should not be contaminated by nearby sources, as the selected extraction radius of 2 arcsec is free of other bright sources (see Table \ref{xsources}). No short-term variability is found in the XMM-\textit{Newton} lightcurves.

As a final comment, we would like to point out that it is quite difficult to find a physical origin for the very steep powerlaw needed to fit the soft X-ray spectrum of NGC~7674. This is a common problem in the analysis of Compton-thick Seyfert galaxies but, as already suggested by \citet{imf02} and \citet{gua04}, it may be the result of a blending of strong emission lines which mimic a continuum component in low resolution spectra. Indeed, this interpretation turns out to be true in the few objects were high resolution spectra are available, as in NGC~1068 \citep{kin02,brink02}, Circinus \citep{Sambruna01b} and Mrk~3 \citep{bianchi05b,sako00b}. This seems also to be the case for NGC~7674. A fit of equivalent statistical quality ($\chi^2=39/37$ d.o.f.) is achieved by modelling the soft spectrum with a powerlaw with the same photon index of the primary continuum (now $\Gamma\simeq1.85$), plus a number of emission lines, mostly from H- and H-like O, Ne, Mg and Si. Therefore, an interpretation of the soft X-ray spectrum in terms of emission from a gas photoionized by the nuclear continuum is the most likely one also in this object, even if high resolution spectra would be required to finally clarify this issue.

\begin{table}

\caption{\label{7674table}NGC~7674: best fit parameters for the observations analyzed in this paper (see text for details).}

\begin{center}

\begin{tabular}{c|cc}
&\textbf{SAX} & \textbf{XMM} \\
&\textit{(1996)} &\textit{(2004)}\\
\hline
&~ & ~ \\
F$_\mathrm{2-10}$ ($10^{-12}$ erg cm$^{-2}$ s$^{-1}$) & $0.75\pm0.06$ & $0.70\pm0.07$\\
$\Gamma$ & $2.05^{+0.19}_{-0.18}$ & $1.7^{+0.3}_{-0.2}$\\
$\Gamma_s$ & $4.6^{+1.2}_{-0.8}$ & $3.3\pm0.2$\\
EW$_{6.4}$ (keV) & $630^{+200}_{-230}$ & $370^{+160}_{-170}$\\
F$_{6.4}$ ($10^{-6}$ ph s$^{-1}$) & $7.7^{+2.4}_{-2.8}$ & $4.2^{+1.8}_{-1.9}$\\
EW$_{6.97}$ (keV) & $340^{+200}_{-230}$ & $480^{+220}_{-300}$\\
F$_{6.97}$ ($10^{-6}$ ph s$^{-1}$) & $3.7^{+2.1}_{-2.4}$ & $4.8^{+2.2}_{-3.0}$\\
$\chi^2$/dof &  63/59 & 45/44\\
\end{tabular}
\end{center}

\end{table}

\subsection{NGC~4968}

\begin{figure*}
\begin{center}
\epsfig{file=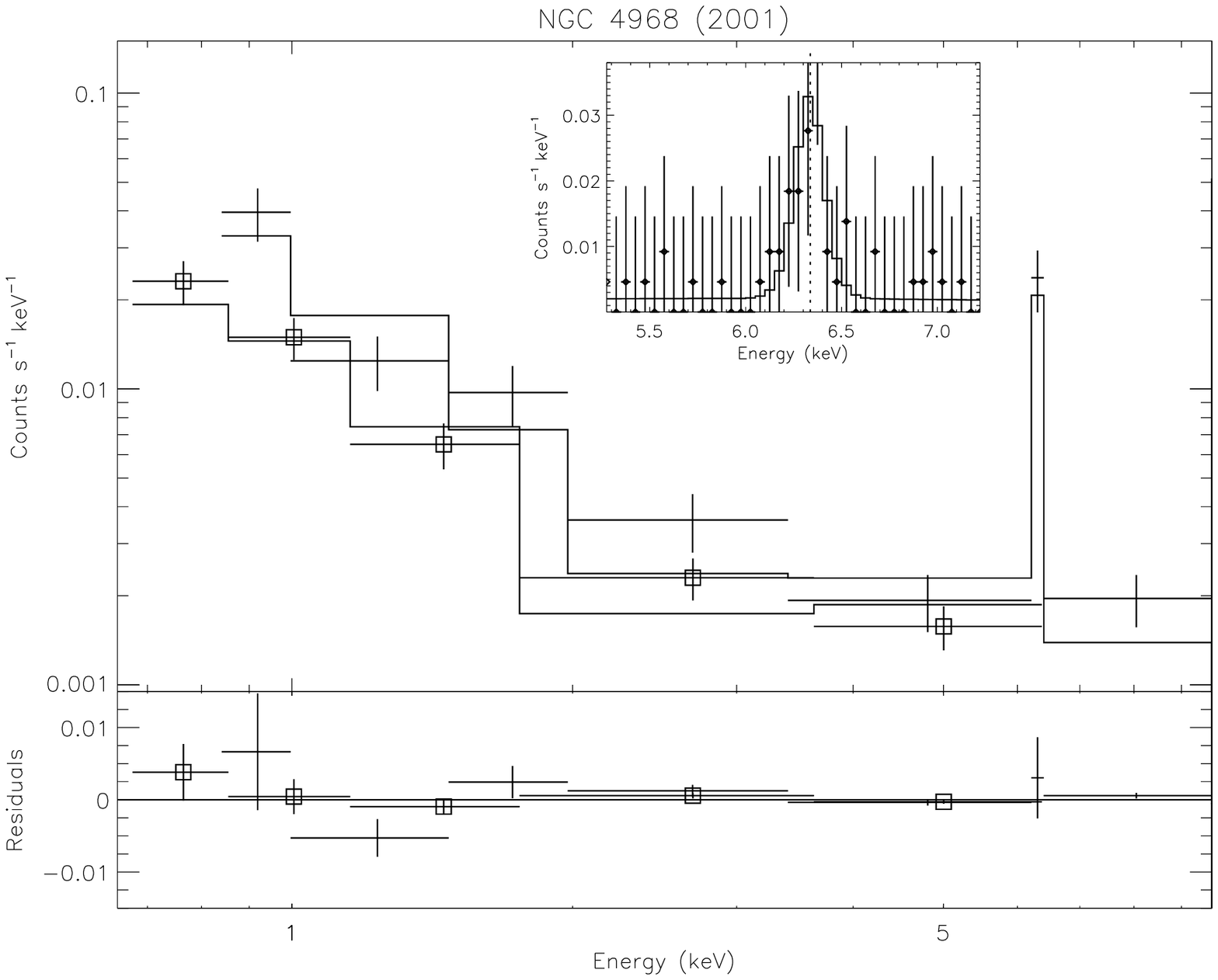, width=8.5cm}
\hspace{0.5cm}
\epsfig{file=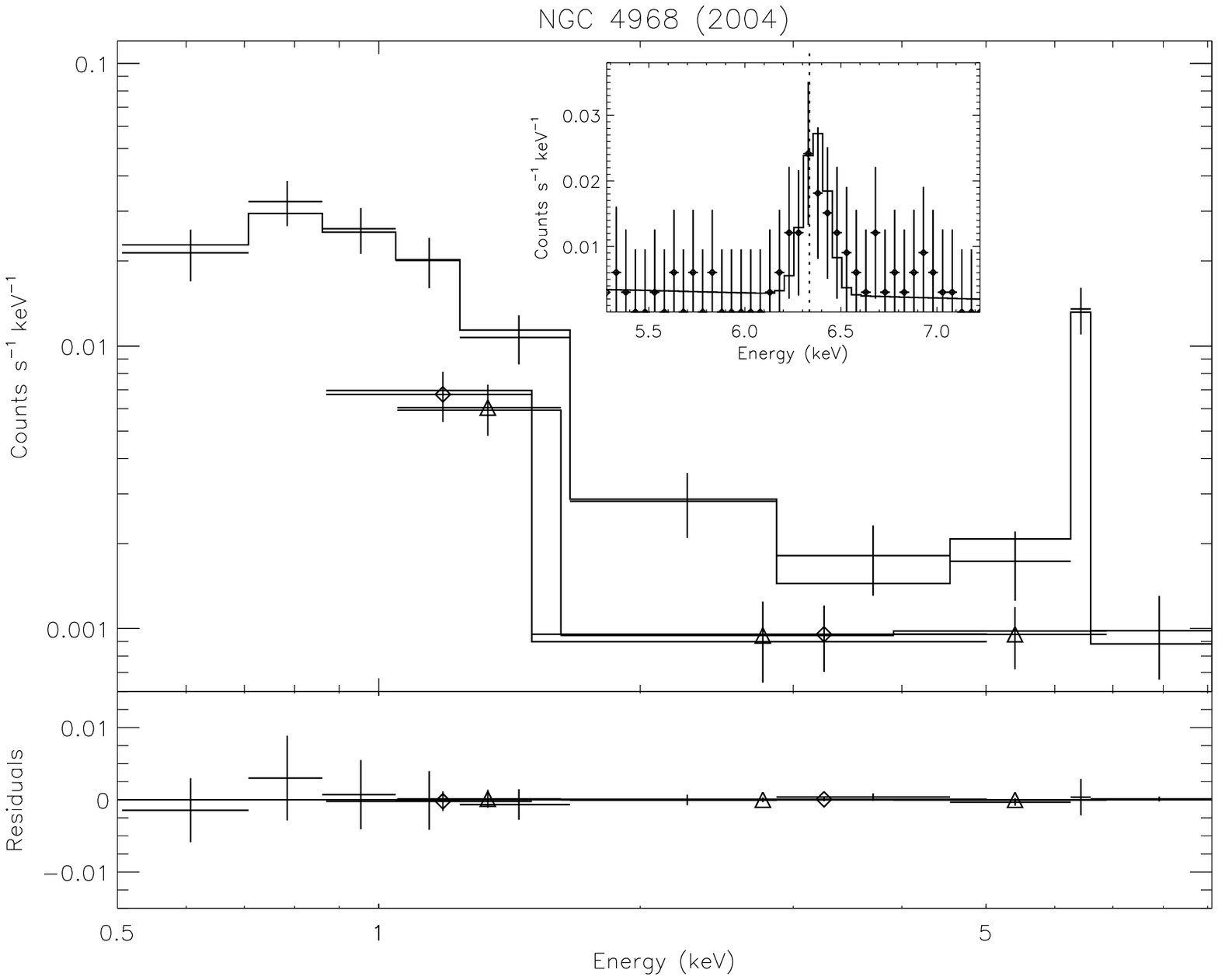, width=8.5cm}
\end{center}
\caption{\label{4968spectrum}NGC~4968: data (\textit{pn}: crosses, \textit{MOS1}: diamonds, \textit{MOS2}: triangles, \textit{Combined MOS}: squares) and best fit model for the first (\textit{left}) and the second (\textit{right}) XMM-\textit{Newton} observation. In the insets: `local fits' on the background-subtracted, unbinned pn spectra around the rest-frame iron line energy, marked with a broken line (see text for details).}
\end{figure*}

NGC~4968 is a Seyfert 2 galaxy which was not detected by \textit{GINGA} \citep{awakiphd} or in the RASS, with an upper limit on the PSPC count rate of 0.03, corresponding roughly to a 0.5-2 keV flux of $10^{-13}$ erg cm$^{-2}$ s$^{-1}$, assuming a $\Gamma=1.7$ powerlaw spectrum \citep{rush96}. A 2-10 keV flux of a few $10^{-13}$ erg cm$^{-2}$ s$^{-1}$ and a poor spectrum consistent with various models characterized the \textit{ASCA} observation \citep{turner97}. Preliminary results on a 2001 XMM-\textit{Newton} observation were presented by \citet{matt02c}, clearly showing that the source was reflection-dominated.
We present here full results on this XMM-\textit{Newton} observation, together with a later one. The two spectra are fully compatible each other and are well fitted by the same model adopted for NGC~7674. The only emission line apparent in the data is the neutral Fe K$\alpha$ line: Table \ref{4968table} and Fig. \ref{4968spectrum} show the best fit parameters and the spectra. No significant variability is found in fluxes or spectral shape between the two observations. Moreover, the XMM-\textit{Newton} lightcurves of each observation are consistent with a constant source.

\begin{table}

\caption{\label{4968table}NGC~4968: best fit parameters for the observations analyzed in this paper (see text for details).}

\begin{center}

\begin{tabular}{c|cc}
&\textbf{XMM1} & \textbf{XMM2} \\
&\textit{(2001)} & \textit{(2004)}\\
\hline
&~ & ~ \\
F$_\mathrm{2-10}$ ($10^{-12}$ erg cm$^{-2}$ s$^{-1}$) & $0.27\pm0.08$ & $0.23\pm0.08$\\
$\Gamma$ & 1.7$^*$ & 1.7$^*$ \\
$\Gamma_s$ & $3.4^{+0.7}_{-0.6}$ & $2.7\pm0.3$\\
EW$_{6.4}$ (eV) & $1900\pm900$ & $3200\pm1100$\\
F$_{6.4}$ ($10^{-6}$ ph s$^{-1}$) & $7.1^{+3.1}_{-3.4}$ & $7.3^{+2.4}_{-2.5}$\\
$\chi^2$/dof & 17/10 & 9/10\\
\end{tabular}
\end{center}

$^*$ Fixed
\end{table}

\subsection{\label{iras}IRAS~13218+0552}

The source is the only object in our sample with a moderately large redshift and classified as a Type 1. After the HEAO A-1 detection, the first X-ray observation of IRAS~13218+0552 was performed on July 2000 by BeppoSAX, which detected a source with a flux more than 60 times fainter and a low quality spectrum: N$_H$ and $\Gamma$ are poorly constrained, but the spectrum seems to be steep ($\Gamma\sim$2). Interestingly enough, there seems to be an excess flux in the PDS, suggesting moderately Compton--thick absorption; however, at these count-rate levels ($0.08\pm0.05$ counts s$^{-1}$) confusion is a serious issue and we cannot draw any definite conclusion.

\begin{figure}
\epsfig{file=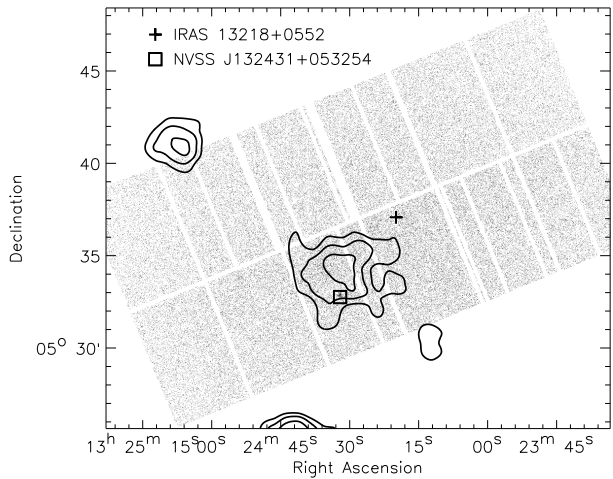, width=8cm}
\caption{\label{irassaxcontour}IRAS~13218+0552: BeppoSAX contour levels superimposed over the EPIC pn field. The target source and a nearby object are marked with a cross and a square, respectively.}
\end{figure}

However, the XMM-\textit{Newton} observation failed to detect IRAS~13218+0552, with an upper limit of $3.9\times10^{-14}$ erg cm$^{-2}$ s$^{-1}$ in the EPIC pn. The closest bright source, located at $\simeq5$ arcmin from the target, corresponds to the radio source NVSS~J132431+053254 (see Table \ref{xsources}). The X-ray spectrum of this source is fully consistent with that of the source observed by BeppoSAX (an unabsorbed powerlaw with $\Gamma\simeq2$), even if its flux is significantly lower, being $5.3\times10^{-14}$ erg cm$^{-2}$ s$^{-1}$. We show in Fig. \ref{irassaxcontour} the contour plot of the MECS observation superimposed on the EPIC pn image, marking the positions of IRAS~13218+0552 and NVSS~J132431+053254: it seems likely that the source observed by BeppoSAX is the latter, so that the target object was undetected also in this observation. Moreover, in a \textit{ROSAT} observation of the same field, no source is apparent at the coordinates of IRAS~13218+0552.

\subsection{NGC~1667}

After the X-ray detection by HEAO A-1 at a flux of $1.9\times10^{-11}$ erg cm$^{-2}$ s$^{-1}$, the source was not detected by \textit{GINGA} and presented a very low flux in the \textit{ASCA} observation \citep{turner97,pappa01}. Indeed, our re-analysis of the \textit{ASCA} data shows that the source is practically undetected above 2 keV. Moreover, only an upper limit was found by the PSPC within the RASS \citep{rush96}.

\begin{figure}
\epsfig{file=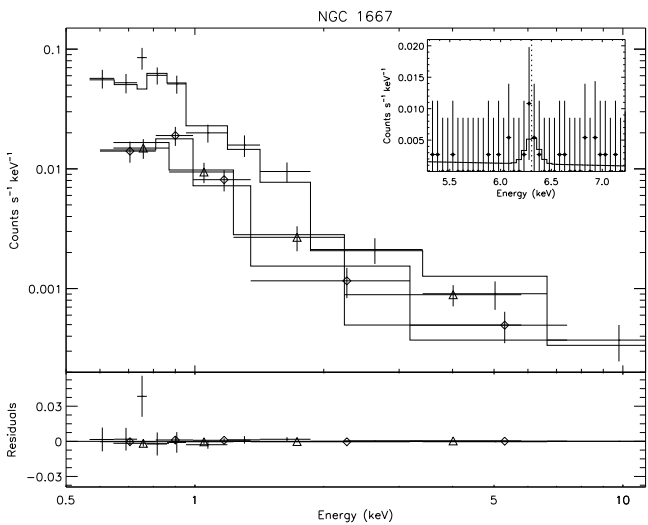, width=8.5cm}
\caption{\label{1667spectrum}NGC~1667: XMM-\textit{Newton} spectra (\textit{pn}: crosses, \textit{MOS1}: diamonds, \textit{MOS2}: triangles) and best fit model. In the inset: `local fit' on the background-subtracted, unbinned pn spectrum around the rest-frame iron line energy, marked with a broken line (see text for details).}
\end{figure}

A low flux level similar to the one found by \textit{ASCA} characterizes the XMM-\textit{Newton} observation. A fit with a simple powerlaw leads to a very steep photon index ($\Gamma\simeq3$), but with large residuals on the softer and the harder parts of the spectrum, resulting in an unacceptable reduced $\chi^2$, greater than 3. On the other hand, the large ratio between the [OIII] and the far infrared (IR) fluxes, together with a relatively low X-ray flux, hint to a Compton-thick source (see Tables \ref{sample}, \ref{diagrams} and Sect. \ref{LMA} for details). Therefore, we tried a model with a pure reflection component and a soft excess: now the fit is perfectly acceptable ($\chi^2=17/14$ d.o.f: see Table \ref{1667table} and Fig. \ref{1667spectrum}). The model includes an emission line at $0.87^{+0.03}_{-0.02}$ keV (likely {Ne\,\textsc{ix}} K$\alpha$), required at the 97\% confidence level according to F-test. The presence of a strong iron line is suggested by the `local fit' (see inset in Fig. \ref{1667spectrum}), with a flux of $1.0^{+1.4}_{-0.6}\times10^{-6}$ ph s$^{-1}$, consistent with the upper limit found in the global fit. This would correspond to an EW of $\simeq600$ eV. The observed 2-10 keV flux is $1\times10^{-13}$ erg cm$^{-2}$ s$^{-1}$ and no significant short-term variability trends are present in the lightcurve.

\begin{table}
\caption{\label{1667table}NGC~1667: best fit parameters for the observation analyzed in this paper (see text for details).}
\begin{center}
\begin{tabular}{c|c}
& \textbf{XMM}\\
\hline
&~ \\
F$_\mathrm{2-10}$ ($10^{-12}$ erg cm$^{-2}$ s$^{-1}$) & $0.10\pm0.03$\\
$\Gamma$ & 1.7$^*$ \\
$\Gamma_s$ & $3.2^{+0.4}_{-0.3}$ \\
EW$_{6.4}$ (eV) & $<600$ \\
$\chi^2$/dof & 17/14 \\
\end{tabular}
\end{center}

$^*$ Fixed
\end{table}

\section{Discussion}

\begin{table*}
\caption{\label{fluxes}The X-ray history of the four sources in our sample.}
\begin{center}
\begin{tabular}{cccccc}
\hline
&&&&&\\
& \textbf{HEAO A-1} & \textbf{GINGA} & \textbf{ASCA} & \textbf{BeppoSAX} & \textbf{XMM-Newton}\\
& Flux \textit{[Date]}&Flux \textit{[Date]}&Flux \textit{[Date]}&Flux \textit{[Date]}&Flux \textit{[Date]}\\
&&&&&\\
\hline
&&&&&\\
NGC~7674 & $24\pm5$ \textit{[1977/78]}$^a$ & $8\pm2$ \textit{[1989]}$^b$ & -- & $0.75\pm0.06$ \textit{[1996]}$^d$& $0.70\pm0.07$ \textit{[2004]}$^d$\\
&&&&&\\
\multirow{2}*{NGC~4968} & \multirow{2}*{$22\pm3$ \textit{[1977/78]}$^a$} & \multirow{2}*{$<5$ \textit{[1989]}$^b$} & \multirow{2}*{$0.2^{+0.4}_{-0.1}$ \textit{[1994]}$^d$} & \multirow{2}*{--} & $0.27\pm0.08$ \textit{[2001]}$^d$\\
&&&&& $0.23\pm0.08$ \textit{[2004]}$^d$\\
&&&&&\\
IRAS~13218+0552 & $18\pm5$ \textit{[1977/78]}$^a$ & -- & -- & $0.28\pm0.06$ $^*$ \textit{[2000]}$^d$ & $<0.039$ \textit{[2004]}$^d$\\
&&&&&\\
\multirow{2}*{NGC~1667} & \multirow{2}*{$19\pm4$ \textit{[1977/78]}$^a$} & $<7$ \textit{[1988]}$^c$ & \multirow{2}*{$<0.15$ \textit{[1994]}$^d$} & \multirow{2}*{--} & \multirow{2}*{$0.10\pm0.03$ \textit{[2004]}$^d$}\\
&& $<3$ \textit{[1990]}$^c$ &&&\\

\end{tabular}
\end{center}
Fluxes are in 10$^{-12}$ erg cm$^{-2}$ s$^{-1}$ (2-10 keV). $^*$ This flux is likely that of a nearby source, NVSS~J132431+053254 (see text for details).

\textit{References}: $^a$ Based on the HEAO A-1 count rates and conversion factor given in \citet{wood84}; $^b$ \citet{awaki91} $^c$ \citet{polletta96} and references therein, but see also Sect. \ref{ginga} $^d$ This paper

\end{table*}

\subsection{Classification of the objects in the sample}

\begin{figure*}
\begin{center}
\epsfig{file=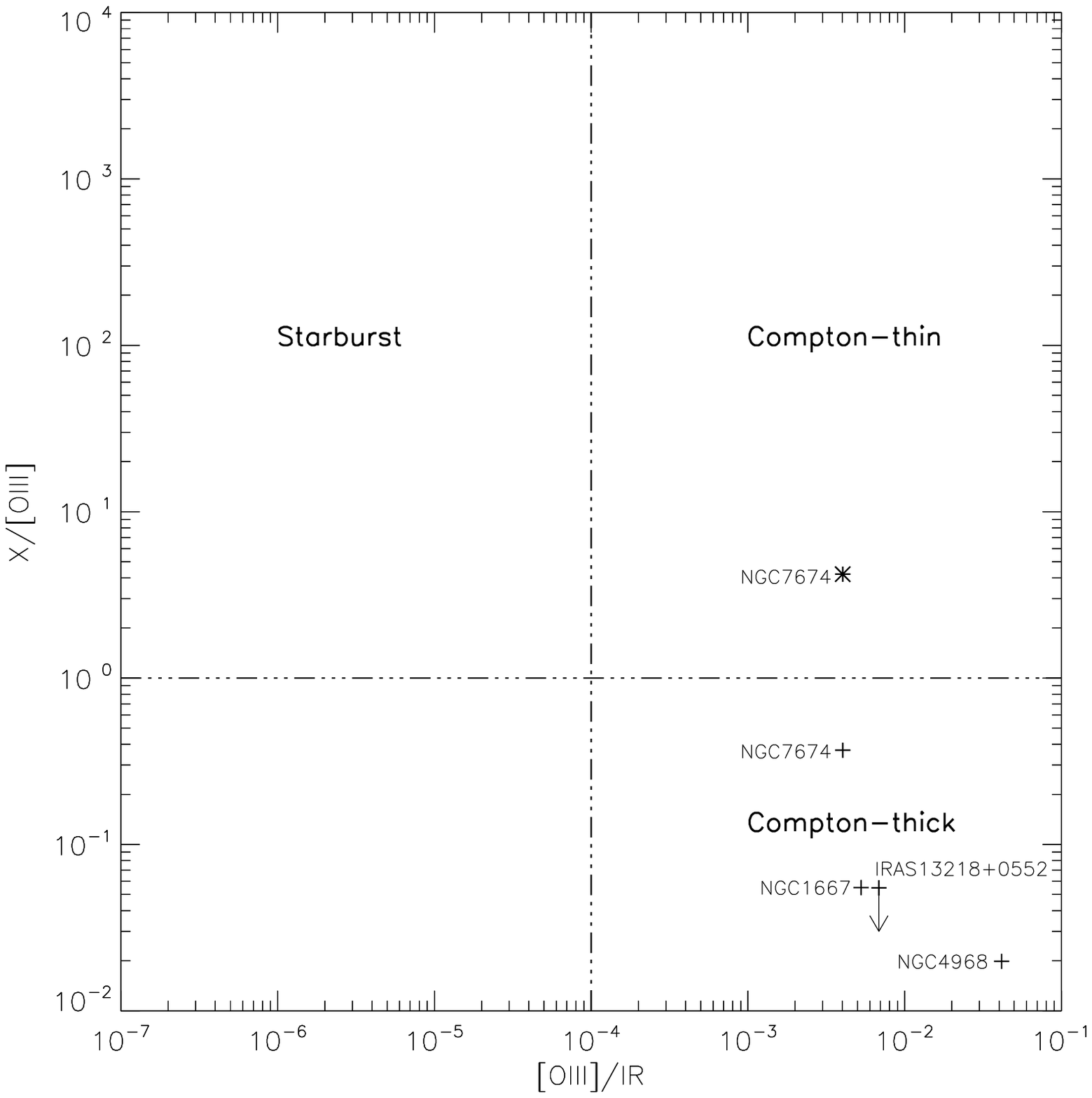, width=8cm,height=7cm}
\hspace{0.5cm}
\epsfig{file=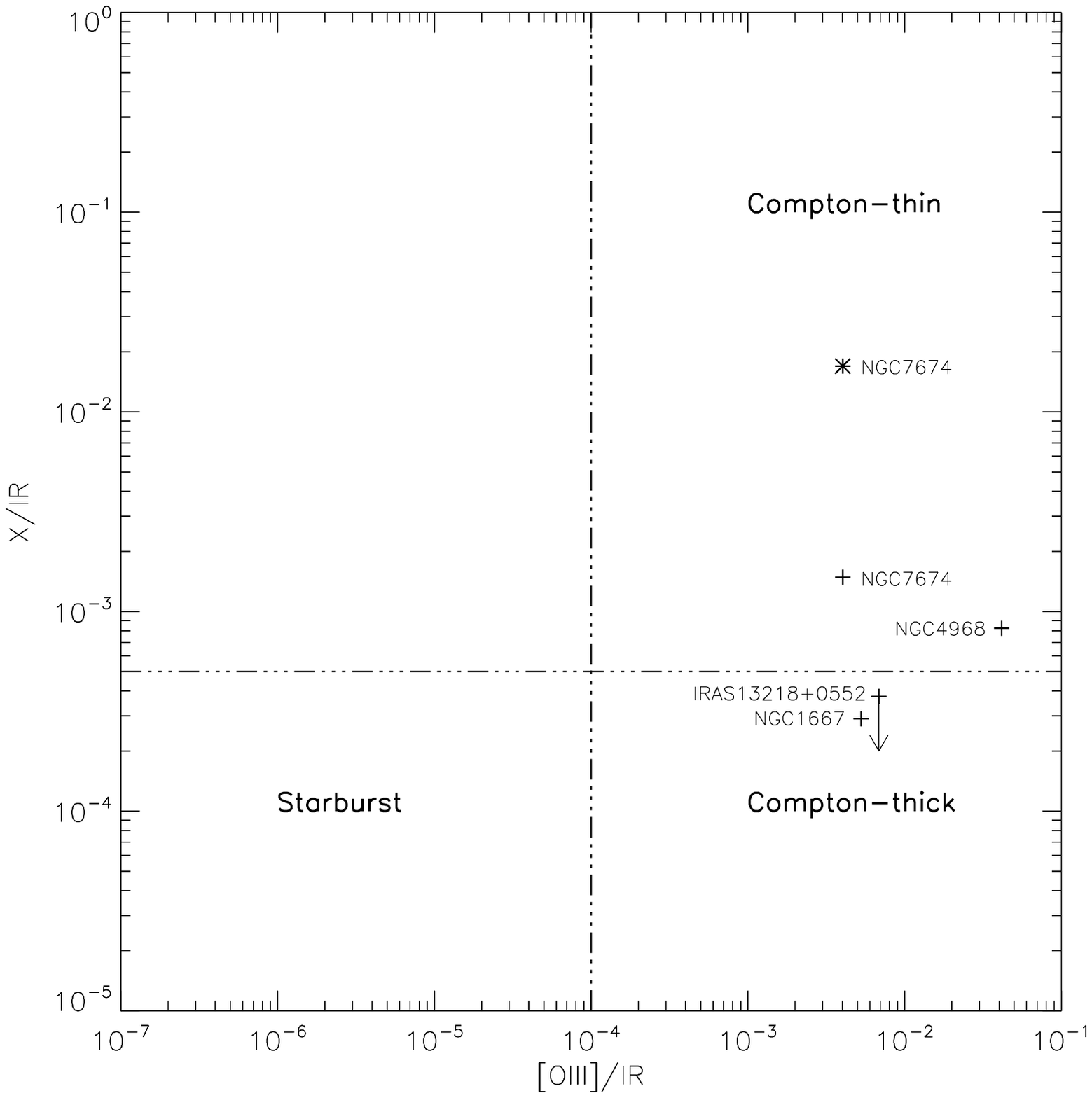, width=8cm,height=7cm}
\end{center}
\caption{\label{diagrams}F$_X$/F$_{IR}$, F$_{[OIII]}$/F$_{IR}$ and F$_X$/F$_{[OIII]}$ diagrams after \citet{pb02}. See Tables \ref{sample} and \ref{diagrams} for data (the X-ray fluxes are taken from the XMM-\textit{Newton} observations, except for NGC~7674, whose \textit{GINGA} flux is also used and the relative point is marked by a star instead of a cross).}
\end{figure*}

All the sources included in our sample turn out to be likely Compton-thick. While the spectral analysis presented in this paper strongly favour this scenario at least for NGC~7674 and NGC~4968 (see Sect. \ref{analysis}), another test can be done on the basis of their [OIII], infrared (IR) and X-ray fluxes, as explained, for example, by \citet{pb02}.  Looking at the diagrams shown in Fig. \ref{diagrams}, it is clear that all the four sources populate or are very close to the regions of the Compton-thick Seyfert galaxies. However, we note here that it should be better to define them reflection-dominated objects, since these diagrams cannot distinguish between highly obscured and `switched-off' sources. Indeed, in at least one case, NGC~7674, the diagrams in Fig. \ref{diagrams} show that the source moved from the Compton-thin to the Compton-thick region between two X-ray observations, thus being a good `switched off' candidate. We will discuss this object further in Sect. \ref{7674off}.

\subsection{\label{LMA}A re-assessment of LMA identifications}

None of the sources in our sample was caught in a high state, comparable to the one measured by HEAO A-1. Therefore, before speculating on large flux variations, we should first re-assess the reliability of the LMA fluxes.

\subsubsection{Wrong identifications}

\begin{figure*}
\begin{center}
\epsfig{file=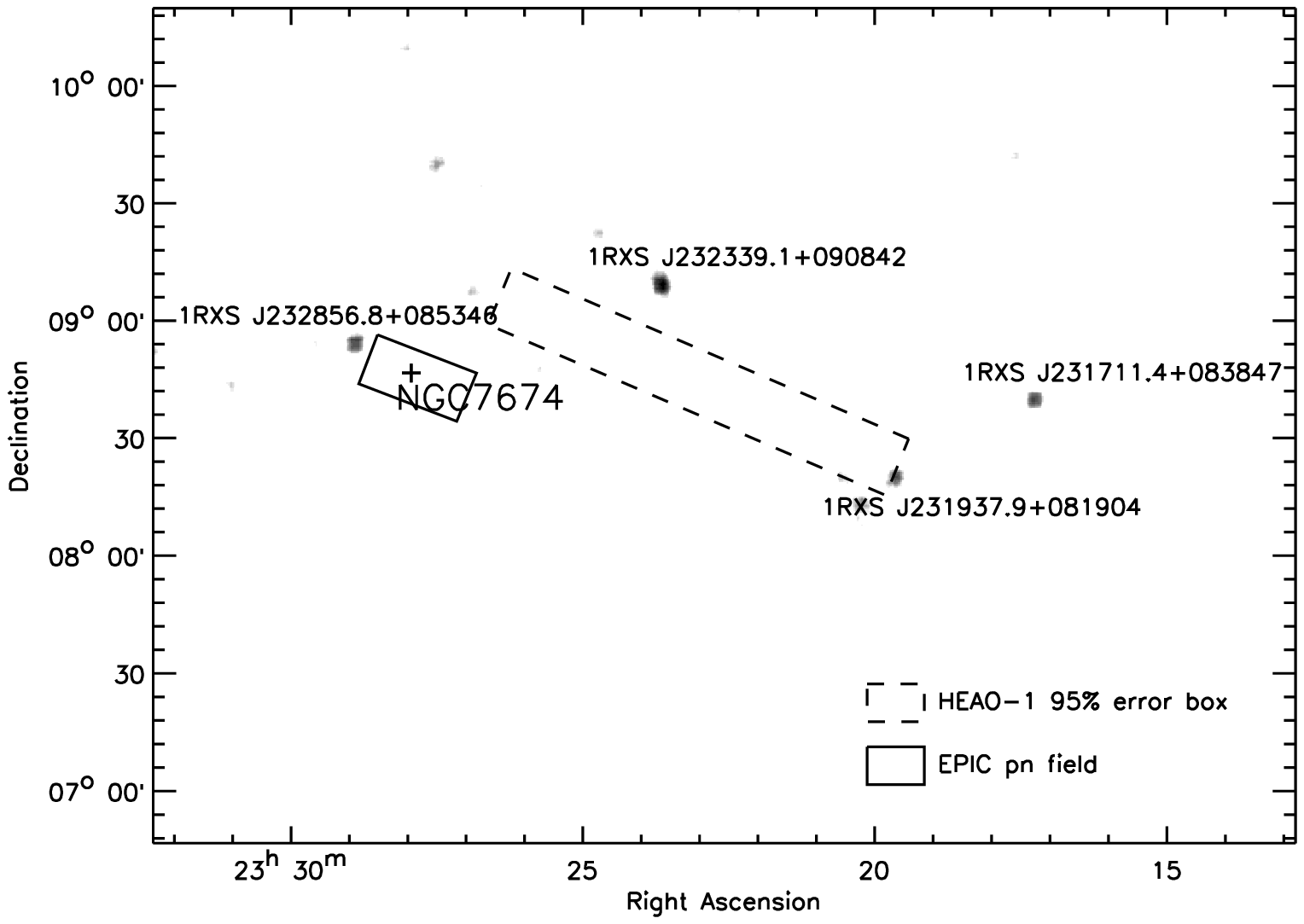, width=8cm}
\hspace{0.5cm}
\epsfig{file=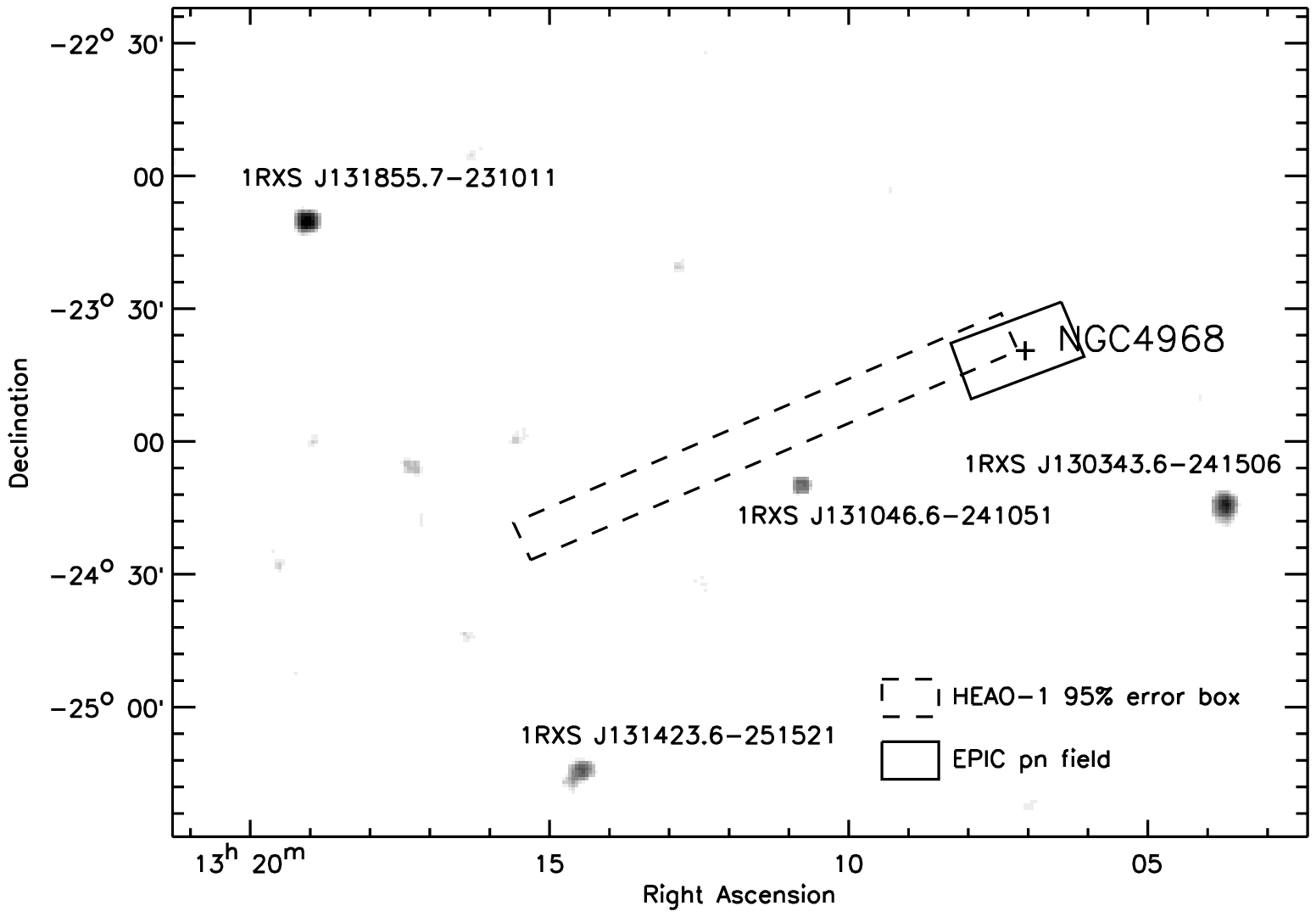, width=8.1cm}

\vspace{1cm}

\epsfig{file=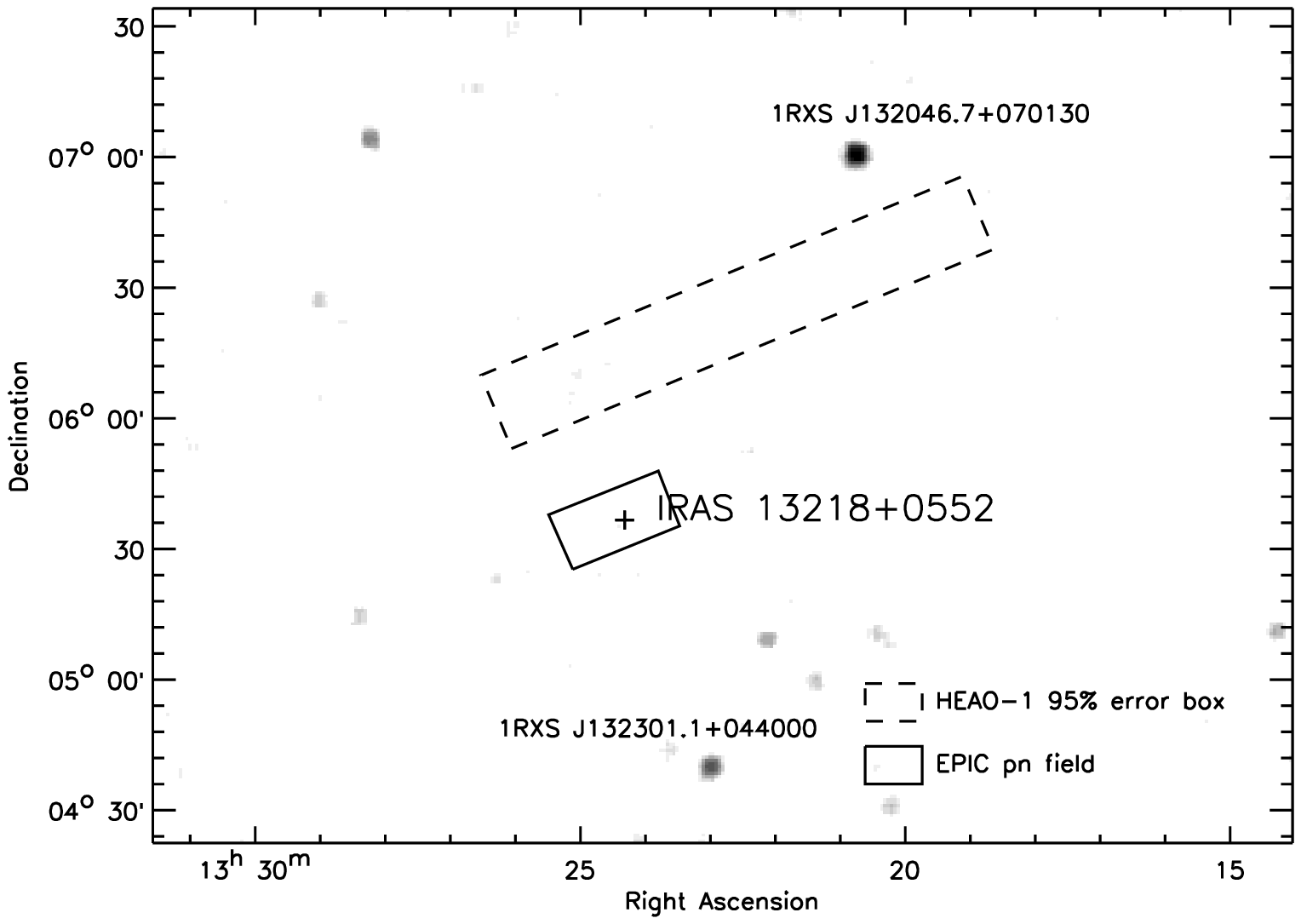, width=8cm}
\hspace{0.5cm}
\epsfig{file=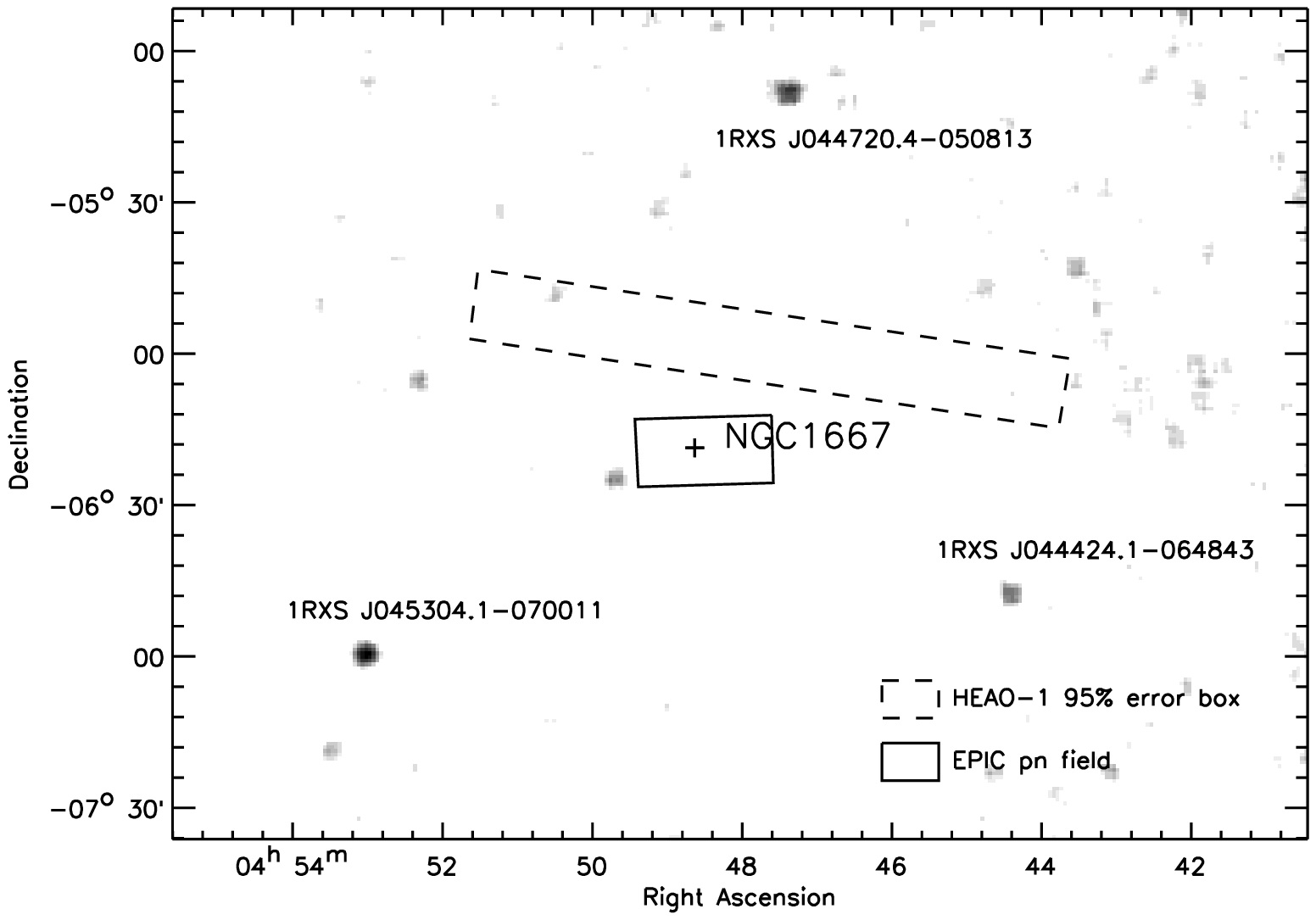, width=8.1cm}
\end{center}
\caption{\label{heaofields}HEAO A-1 95\% error boxes and EPIC pn fields of the XMM-\textit{Newton} observations analyzed in this paper overplotted on the RASS images. A cross points to the coordinates of the LMA counterpart, while the brightest RASS sources are labelled (see text for details).}
\end{figure*}

A first test is to check the optical identifications given by \citet{grossan92} for the four sources in our sample. We reconstructed the rectangular HEAO A-1 error boxes using the coordinates reported by \citet{wood84} and overplotted the EPIC pn fields together with the position of the AGN which constitute our sample. The results are shown in Fig. \ref{heaofields}. With the only exception of NGC~4968, all the sources fall largely outside of the 95\% LASS error box. Note that the actual error region is much smaller than the latter, because the LMA identifications also take into consideration the MC pointings. One of the sources, IRAS~13218+0552, was part of a 22-objects subsample of the LMA analyzed in detail by \citet{rem93}, who included a figure very similar to the one presented here. Nonetheless, he suggested that in this case (and in some others) the LASS error boxes should be considered as $\simeq50\%$ contours. However, it is clear that these identifications are much less robust than for other LMA sources.

If this is the case, we should also look for contamination by nearby sources, which could contribute to the total flux measured by HEAO A-1. The LASS error boxes shown in Fig. \ref{heaofields} are overplotted on the RASS images of the same fields. No bright source lies inside any of the HEAO A-1 boxes. Moreover, we measured fluxes of the brightest objects present in the EPIC fields of each target source. The results are shown in Table \ref{xsources}: none of these sources have X-ray fluxes comparable to the one measured by the LMA and are all (with the exception of NVSS~J132431+053254 and SDSS~J132442.44+052438.9 in the otherwise empty field of IRAS~13218+0552), much dimmer than the AGN. Even if some of them can be highly variable sources, we have no evidence that their fluxes could have significantly contaminated the LASS count rates.

\begin{table*}
\caption{\label{xsources}Main data on the brightest sources in the EPIC fields analysed in this paper (see text for details).}
\begin{center}
\begin{tabular}{cccccc}
\textbf{Source} & \textbf{Identification}$^a$ & \textbf{RA DEC (J2000)} & \textbf{Distance}$^b$ & \textbf{F$_\mathrm{0.5-2 keV}$}$^c$ & \textbf{F$_\mathrm{2-10 keV}$}$^c$\\
\hline
~&~&~&~&~&~\\
7674-X1 & XMMU J232757.4+084744 $^1$ & 23$^h$27$^m$57$^s$.4 +08\degr47\arcmin44\arcsec & 1.1 & 0.62 & 0.22 \\
7674-X2 & 1WGA J2328.1+0849 $^2$ & 23$^h$28$^m$07$^s$.3 +08\degr49\arcmin07\arcsec & 3.6 & 0.56 & 1.4 \\
7674-X3 & XMMU J232728.0+085206 $^1$ & 23$^h$27$^m$28$^s$.0 +08\degr52\arcmin06\arcsec & 7.1 & 0.71 & 0.28\\
7674-X4 & 1WGA J2327.4+0845 $^2$ & 23$^h$27$^m$26$^s$.7 +08\degr45\arcmin11\arcsec & 7.6 & 1.4 & 0.035\\
7674-X5 & 2MASX J23282907+0853013 $^3$ & 23$^h$28$^m$28$^s$.9 +08\degr52\arcmin58\arcsec & 10.2 & 1.6 & 1.4 \\
~&~&~&~&~&~\\
IRAS-X1 & NVSS~J132431+053254 $^4$ & 13$^h$24$^m$32$^s$.0 +05\degr32\arcmin51\arcsec & 5.2 & 0.48 & 0.53\\
IRAS-X2 & SDSS~J132442.44+052438.9 $^5$ & 13$^h$24$^m$42$^s$.7 +05\degr24\arcmin37\arcsec & 13.7 & 4.8 & 12.2\\
~&~&~&~&~&~\\
1667-X1 & XMMU J044835.7-061810 $^1$ & 04$^h$48$^m$35$^s$.7 -06\degr18\arcmin11\arcsec & 0.95 & 0.21 & 0.11\\
1667-X2 & XMMU J044833.1-061716 $^1$ & 04$^h$48$^m$33$^s$.2 -06\degr17\arcmin16\arcsec & 2.1 & 0.23 & 0.45\\
1667-X3 & XMMU J044851.4-062135 $^1$ & 04$^h$48$^m$51$^s$.4 -06\degr21\arcmin35\arcsec & 4.4 & 0.13 & 0.18\\
~&~&~&~&~&~\\
\hline
\end{tabular}
\end{center}
$^a$ References: (1) This paper - (2) \citet{wgacat} - (3) \citet{2masscat} - (4) \citet{nraosur} - (5) \citet{stra03} $^b$ Distance (in arcmin) of the source from the target - $^c$ In units of $10^{-13}$ erg cm$^{-2}$ s$^{-1}$
\end{table*}

\subsubsection{Wrong fluxes}

Even if the optical identification given by \citet{grossan92} are very uncertain, there is no clear evidence of contamination from other sources. Therefore, it is possible that the LMA fluxes are wrong. Indeed, there are two reasons to suspect that. The first one is that the fluxes of these sources are among the lowest in the sample and, in the cases of IRAS~13218+0552 and NGC~1667, they are just above the flux limit of the sample itself, with errors around 20$\%$. A second issue refers to the fact that HEAO A-1, due to an hardware failure, lost all its spectral resolution, so that it should be considered as a `large X-ray photometer over a range of 1-20 keV', to use the words of \citet{grossan92}. The measured count rates were then transformed to a flux density at 5 keV adopting an empirical flux conversion derived from a comparison with the HEAO A-2 instrument and assuming a spectral shape of a powerlaw with a photon index of 1.7. The last assumption gives reasonable results for unabsorbed or even moderately absorbed AGN, but can lead to completely wrong fluxes in Compton-thick sources. If the primary continuum is blocked up to 10 keV or higher energies, the measured LASS count rates are totally unrelated to the 2-10 keV flux of the source, because they refer, in an unknown way, to the total band of the instrument, up to 20 keV.

We performed a rough test with \textsc{Xspec} to see if this scenario is consistent with the measured PDS fluxes in the two sources observed by BeppoSAX. We first calculated a 1-20 keV count-rate (arbitrarily normalized) which reproduced the 2-10 keV HEAO flux with a model consisting of an unabsorbed powerlaw with $\Gamma=1.7$. Then, we produced a model consisting of a bare Compton-reflection component (with a 2-10 keV flux like the one observed by XMM-\textit{Newton}) and a highly obscured (N$_\mathrm{H}$ of several $10^{24}$ cm$^{-2}$) powerlaw, yielding the same count-rate. The resulting 10-20 keV flux is, for both NGC~7674 and IRAS~13218+0552, a factor $\simeq100$ larger than the one observed by BeppoSAX, showing that, at least in these objects, it is unlikely that the count-rate observed by HEAO A-1 can be interpreted in terms of high energy emission of an otherwise obscured object. We note that our test assumes a flat response of the detector in the whole band 1-20 keV. In a more realistic case, where the instrumental effective area decreases at high energies, the difference between the expected flux and the one observed by the PDS would be larger.

In conclusion, even if we do not find any conclusive evidence against the identification or the flux of the sources in our sample, the above-mentioned issues may affect, in principle, all the low-flux objects in the \citet{grossan92} catalog, which should be treated with due caution. However, it is worthwhile stressing that the criterion adopted in our sample clearly puts a strong bias towards potentially fake HEAO A-1 identifications or detections, since all the targets were not confirmed at the same flux level by other observations. For our purposes, all the preceding discussion does not allow us to speculate on large X-ray variations of the sources in our sample on the basis of their HEAO A-1 fluxes only. In particular, NGC~4968 is the only object which lies reasonably close to the LMA error box, but, apart from its LMA flux, it does not present any significant variation for several years, similarly to NGC~1667 (see Table \ref{fluxes}). On the other hand, if we relax the constraints given by the HEAO error boxes, as suggested by \citet{rem93}, in one case, NGC~7674, the LMA flux could be in principle suggestive of a past high flux state of the source, considering the following \textit{GINGA} flux, which is significantly larger than the BeppoSAX and the XMM-\textit{Newton} ones. Finally, another object, IRAS~13218+0552, is not detected at all by XMM-\textit{Newton}. These two sources deserve some further comments.

\subsection{\label{7674off}NGC~7674: a switched-off source?}

Even not taking into account the HEAO A-1 observation, the source underwent a flux loss by a factor $>\sim10$ in the 7 years separating the \textit{GINGA} and BeppoSAX observation, then remained fairly stable after other 8 years, when the XMM-\textit{Newton} observation was performed (see Table \ref{fluxes}). Therefore, NGC~7674 may well represent a good candidate for being a switched-off object.

\begin{figure}
\epsfig{file=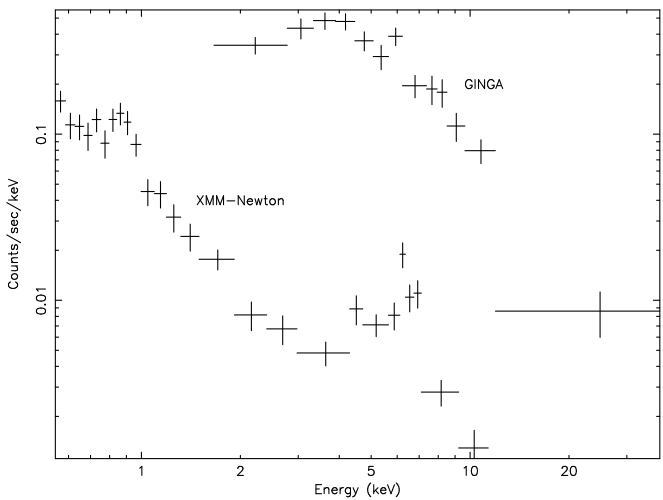, width=6.5cm, angle=-90}
\caption{\label{7674ginga2xmm}NGC~7674: a comparison between the \textit{GINGA} (upper) and XMM-\textit{Newton} (lower) spectra, clearly showing the transition between a transition- to a reprocessing-dominated state (see text for details).}
\end{figure}

This is also supported by a spectral transition between the two flux states. Indeed, the \textit{GINGA } spectrum was a featureless powerlaw, with only an upper limit to the EW of the iron line (80 eV) and an unconstrained value for the neutral absorbing column density. On the other hand, the BeppoSAX and XMM-\textit{Newton} spectral analysis clearly show that the source is actually reprocessing-dominated, with a bare Compton reflection continuum at high energies, an iron line with a EW$\simeq400$ eV and a soft excess at lower energies (see Fig. \ref{7674ginga2xmm}). Moreover, the source lies in the Compton-thick region in the diagnostic diagrams based on the X-ray, IR and [OIII] fluxes (see Fig. \ref{diagrams}).

Such a transition between a transmission- to a reprocessing-dominated spectral state would make NGC~7674 the sixth Seyfert 2 galaxy to show this behaviour \citep[see][and references therein]{gua05}. The time elapsed between the \textit{GINGA}, BeppoSAX observation and XMM-\textit{Newton} observations allows us to make an estimate of the distance of the reflector from the Black Hole (BH). If the primary source switched off shortly after the \textit{GINGA} observation, the detection of a reflection echo 15 years after implies a distance of around 5 pc of the material responsible for it. On the other hand, if, instead, it switched off a little before the BeppoSAX observation, the same flux level measured 8 years later puts a lower limit of almost 3 pc. Since it is impossible to know if in these time intervals the source actually switched on again, care should be taken when considering these lower limits.

Another interesting piece of information comes from the ratio between the normalizations of the primary powerlaw in the Compton-thin state and the Compton reflection component in the reprocessing-dominated state, which leads to an estimate of the covering factor of the reflecting matter. In the case of NGC~7674, this value is approximately $R=1.5$, i.e. the subtended solid angle is around $3\pi$. We have calculated the same ratio for all the `changing-look' Seyfert 2 sources in the literature and found values ranging from 0.8 to 1.8 \citep[except for the extreme case of NGC~6300, but in this case the primary continuum is very variable and it is difficult to calculate R,][]{gua02b}. It should be noted that these values must be considered as estimates of the true covering factor, because they are calculated assuming transmission and reflection component fluxes which were not measured simultaneously, thus not necessarily directly linked one to the other. Since in `changing-look' sources the Compton-thick material clearly does not intercept the line of sight, these values of \textit{R} should be compared to the ones found in Seyfert 1s, which are in the range $\simeq0.5-1$ \citep[see e.g.][]{per02,bianchi04}: this is suggestive that the same material \citep[likely the torus envisaged in Unification Models,][]{antonucci93} is responsible for the reflection components observed in the two classes of objects.

\subsection{A misclassified Compton-thick object: IRAS~13218+0552}

While the HEAO A-1 flux is likely the result of a misidentification and the BeppoSAX flux also probably belongs to a nearby source (see previous section), the non-detection of IRAS~13218+0552 by XMM-\textit{Newton} still makes this source peculiar. Indeed, the upper limit derived for its X-ray flux, together with the IR and [OIII] fluxes, clearly puts this object among the Compton-thick sources (see Fig. \ref{diagrams}). This is at odds with its current optical classification. Indeed, the presence of emission lines with components broader than $5\,000$ km s$^{-1}$ and the stellar-like appearance induced \citet{low88} to classify the source as a QSO. However, it was soon clear that IRAS~13218+0552 was peculiar and its extremely red IR continuum won it the name of `reddest known quasar' \citep{low89}. Moreover, \citet{rem93} reported anomalously broad [OIII] emission lines (FWHM $\simeq3\,500$ km s$^{-1}$), likely the result of some substructure in their profile. An asymmetric [OIII] profile, due to a blueshifted component, was also reported by \citet{zheng02}.

The \textit{HST} image in the R band is strongly suggestive of a merger between two galaxies in its latest stage, according to \citet{boyce96}. They also noted that, given its low nuclear luminosity, it should be strictly classified as a Seyfert 1, instead of a QSO, but its colors clearly points toward a very large obscuration, so that it is very likely that it hosts a buried QSO. An extreme velocity outflow (EVOF) component was finally unambiguously found in the H$\beta$ and [OIII] line profiles, with velocities of the order of $2\,000$ km s$^{-1}$ \citep{lip03}. This detection, together with the previous results, led these authors to conclude that IRAS~13218+0552 is likely the result of a recent merging process and the nuclear energy is due to the composite activity of a hidden QSO and a starburst, the latter implied also by the tentative detection of Wolf-Rayet features. Indeed, the starburst component could even be the dominant one, as suggested by the optical emission line ratios reported by \citet{kvs98}\footnote{The previous classification as a Type 1 object prevented these authors to use, in the case of IRAS~13218+0552, the diagnostic diagrams plotted in Fig. 2 of their paper. However, once the presence of broad lines is excluded, we are allowed to use these line ratios to discriminate between a starburst- or AGN-dominated object.}.

In conclusion, the original classification as a QSO/Seyfert 1 was probably the result of a misinterpretation of the EVOF component as emission from the Broad Line Region. Therefore, the source should be re-classified as a Type 2 object, Compton-thick in the X-rays and likely surrounded by a massive starburst. Within this scenario, it is possible that the PDS excess detected in the BeppoSAX observation (whose MECS spectrum is instead likely dominated by a nearby source: see Sect. \ref{iras}) refers to the nuclear emission of the buried AGN, which pierces through a very large gas column density. However, if an AGN is actively heating the circumnuclear dust, an infrared compact source should be seen with the high spatial resolution imaging provided by \textit{HST} at 1.6-2.2 $\mu$m, as seen in similar Compton-thick nearby Seyfert galaxies such as NGC 1068 \citep{thom01}.

\section{Conclusions}

We have selected a sample of 4 AGN, included in the \citet{grossan92} catalog, which showed in subsequent observations a flux much lower than the one measured with HEAO A-1, thus being good candidates for being `changing-look' sources. None of the sources was caught in a high flux state during the XMM-\textit{Newton} observations. We have shown that, for all the sources, potential problems with the HEAO A-1 source identification and flux measurement prevent us to be certain that the HEAO A-1 data represent a putative `high' state for these objects.

However, based on the high flux state of its \textit{GINGA} observation, a factor of ten higher than in the BeppoSAX and XMM-\textit{Newton} observation, NGC~7674 represents probably the sixth known case of a `changing-look' Seyfert 2 galaxy. This is also supported by a spectral transition between a transmission- to reprocessing-dominated state between the two observations. From the X-ray variability pattern, we can estimate a lower limit of a few parsec to the distance of the reflecting material.

Finally, one of the sources, IRAS~13218+0552, was not detected by XMM-\textit{Newton}, despite being currently classified as a Seyfert 1 with a large [OIII] flux. However, the original classification was likely to be affected by an outflow component in the emission lines. The object likely harbors an highly obscured AGN and should be re-classified as a Type 2 source.

\acknowledgement

We would like to thank the referee, S. Lumsden, for his valuable suggestions. This paper is based on observations obtained with XMM-\textit{Newton}, an ESA science mission with instruments and contributions directly funded by ESA Member States and the USA (NASA).

\bibliographystyle{aa}
\bibliography{sbs}

\end{document}